# Magda – Manager for grid-based data

Wensheng Deng, Torre Wenaus
*Brookhaven National Laboratory, NY 11973, USA*

Magda is a distributed data manager prototype for grid-resident data. It makes use of the MySQL open source relational database, Perl, Java and C++ to provide file cataloging, retrieval and replication services. For data movement, gridFTP, bbftp and scp can be chosen depending on available protocols. Third party transfers are supported. Magda currently catalogs 321K files with total size of 85.8 TB. It was successfully used in the Data Challenge 1 production for the Atlas experiment. It has been used to replicate more than 7 TB of data between BNL and CERN mass stores. We present here the design and implementation of Magda, and how it has been used as a file catalog and replication tool.

## 1. INTRODUCTION

When the Large Hadron Collider starts operation in 2007, it will generate several PetaBytes of data each year. The data and the product from its subsequent processing will be stored and distributed between many major computing centers in Europe, America and Asia. The requirement for the computing power is unprecedented, and is equivalent to hundreds of thousands of today's fastest PC processors, which will be distributed geographically around the world. Physicists must be able to locate data of interest, and send data to where computing power is available. They need convenient data lookup and retrieval tools. That is the motivation for creating Magda [1].

Magda is a distributed data manager prototype for the Atlas experiment. It has been in stable operation since May 2001. The Magda project is supported by the Particle Physics Data Grid (PPDG) [2] project. Magda is an end-to-end application layered over grid middleware. It makes use of the Globus Toolkit [3] wherever applicable.

## 2. DESIGN AND IMPLEMENTATION

### 2.1. Data organization

In the Atlas experiment, data is distributed among globally dispersed storage facilities. In Magda the concept 'site' is used to abstract the storage facility. The concept 'location' is used to denote data locations within a site – typically directory trees. Locations carry attributes characterizing the data stored there, for example whether the data is a master or replica copy. The concept 'host' is used to represent a collection of computers that have access to a defined set of sites. Thus a Magda service having knowledge of the host it is running on is automatically aware of all the data locally accessible to it.

Figure 1 shows the logical relations among site, location and host.

### 2.2. Database Schema

Magda makes use of the MySQL open source relational database, Perl, C++ and Java. The 'core' of the system is a MySQL database. The principal components of the system, implemented as MySQL tables, are:
- File catalog with logical and physical file information and metadata. The file catalog supports the notion of master and replica instances;
- Site, location and host catalogs;
- Logical files can optionally be organized into collections;
- Replication operations are organized into reusable tasks.

### 2.3. Scripts and API's

The database interaction is done via Perl, C++, Java and CGI (Perl) scripts. Scripts and the web interface provide for setting up and managing distributed sites with associated data locations, and the hosts on which data-gathering servers and user applications run; gathering data from the various sorts of data stores; interfacing to users via web interfaces for presenting and querying catalog information and for modifying the system; and replicating and serving files to production and end-user applications.

Perl scripts drive the creation of the MySQL databases making up the system; filling the databases with the metadata describing the system; crawling data sources and filling the databases with catalog content; collecting summary information from the database; cgi presentation of and interaction with the database on the web; file replication; API code auto-generation from the database; user-level tools.

C++, Java and Perl APIs for access to all components of the database are provided. C++ and Java APIs are autogenerated by Perl scripts using the database schema as obtained from MySQL metadata. The interface classes are thus automatically kept in synch with the evolving database. Classes are built with accessors and data members which map directly onto database tables. Types are int, double, or string as appropriate to the DB field type. The interfaces support querying on the database, iterating on and processing returned results, and loading the database. Perl is so wondrously pliable that no code auto-generation is required; generic code provides full access to all databases.

### 2.4. User interface

The web interface supports browsing database components (catalog, sites, locations, hosts), querying the catalog (by date, type, location, filename, owner), and modifying the system (adding and editing sites, locations, hosts, collections, replication tasks).

The command line tools (magda_findfile, magda_putfile, magda_validate and magda_getfile) are provided for users to query the catalog, archive files, validate files and fetch files for use.

### 2.5. Catalog content categorizing

Magda categorizes content in an organizational hierarchy, from highest to lowest level as follows: virtual organization (experiment), group (sub-detector group; physics working group), activity (activity within the group, e.g., test beam), team (e.g., institute participating in activity), personal ('me and my laptop', personal physics analysis).

### 2.6. Logical filename and key

Logical file names are unconstrained by the system. They can be assigned arbitrarily by the user. They are not used to carry attribute information unless the user chooses to use them that way. Default logical name assignment is: 1) the filename for data file; 2) the full path within CVS hierarchy plus filename for source code.

Magda supports associating an arbitrary set of keys with catalog entries, allowing entries to be grouped and selected on via user defined criteria. These are used for example to group simulation files on the basis of the physics, detector coverage, kinematics etc. they contain.

### 2.7. Collection

The catalog is primarily based on manipulation of individual files. Collections of files are supported and may be used where appropriate. There are a number of collection types:
- Logical collections. The logical files contained within the collection are arbitrary and user defined.
- Collections associated with a location. Collection has an associated location, and the members of the collection are the files resident at that location.
- Key collections. The members of the collection are all those logical files with a key matching one associated with the collection.

A given logical file can belong to any number of collections.

### 3. BULK DATA REPLICATION

Automated file replication is supported through the definition of replication tasks which have associated with them:
- A collection of files (logical file names) to be replicated.
- Explicit information on the source from which the files are to be retrieved, including a cache collection if files require staging from mass store.
- Information on the file transport mechanism to be used (currently gridFTP, bbftp or scp).
- Information on the destination location for the replication task, including provision of a destination-side cache if the final location is in mass store.

A set of scripts drive the replication task, obtaining the task details from the database. The most complex case of a task which replicates files from one mass store to another, e.g. CERN stage to BNL rftp, operates with three scripts:

dySource2Cache.pl moves files from the source mass store out into the source-side cache, accessible to the transfer software, checks that adequate disk space is available in the source cache, and cleans out cached files for which the transfer is complete.

dyTransfer.pl moves files over the net from the source-side cache to the destination-side cache.

dyCache2Dest.pl moves files from the destination cache to the final destination location in mass store, and registers the new file location with the replica catalog.

dyTaskManager.pl Manages replication tasks from command line, and allows activation, deactivation, clearing, deleting of tasks.

The interlocks required to manage these asynchronously running steps are done via the database. Each script updates logical file collections associated with the relevant caches to record the progress of the transfer operation and indicate the availability of a given file to the next stage of the transfer. The scripts consult these collections for files they can operate on.

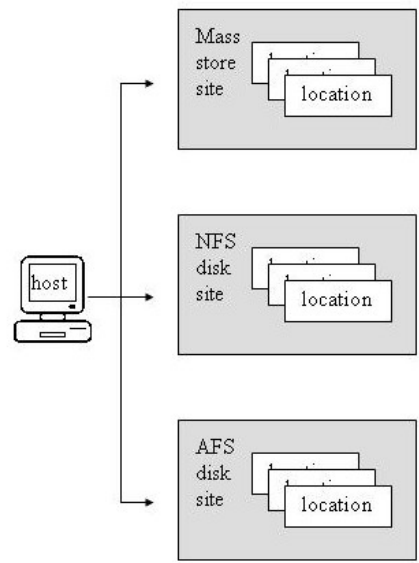

Figure1 Logical relations among site, location and host

## 4. USAGE TO DATE

### 4.1. Atlas Data Challenge 0 and 1

Magda has been successfully used as the Atlas standard data management tool in Atlas Data Challenges 0 and 1 which took place from late 2001 through spring 2003. As of May 2003 there were 321 K files with total size of 85.8 TB registered with Magda. These files are physically located at 18 institutions around the world: University of Alberta in Canada, CERN in Geneva, IN2P3 in Lyon France, CNAF and Milan of INFN in Italy, GridKa at FZK in Germany, IFIC in Spain, IHEP and ITEP in Russia, NorduGrid, RAL in United Kingdom, United States (Boston University, Indiana University, LBNL, University of Chicago, University of Oklahoma, BNL, University of Texas at Arlington). Institutes use Magda to fetch remote files needed as input to local production or for local use. Automatic data replication tasks have copied approximately 7 TB of data between the mass stores of BNL and CERN.

### 4.2. US Atlas grid testbed production

For Atlas DC 1, a series of productions have been done on the US Atlas grid testbed. The job submission and control was managed by GRAT [4], developed at the University of Texas at Arlington. Data management was done with Magda. GRAT is a set of shell and Python scripts which call Magda commands for data moving and cataloging. More than 10 TB of produced data has been replicated to BNL HPSS and cataloged.

### 4.3. Tested on EDG testbed

Magda has been successfully installed and tested on the European Data Grid (EDG) [5] testbed. Installation of Magda and ancillary software was done using pacman [6]. The test distributed Magda tarballs together with a generic application in EDG input sandboxes, sending them to computing elements where they were executed. Test jobs were submitted to five EDG sites. The jobs generated data locally, used magda_putfile to move it from worker nodes to storage locations on EDG Storage Elements, used magda_findfile to check proper file registration, used magda_deletefile to remove files, and magda_findfile again to verify deletion. The summary of this test is available [7].

### 4.4. Other Usage

Using Magda, the RHIC Phenix experiment replicates data from BNL to Stony Brook University, and catalogs data at Stony Brook. Magda is being evaluated by other potential users.

### Acknowledgments

The authors wish to thank the PPDG project and the Atlas experiment for their support and feedback. This work is supported through the Division of High Energy Physics under DOE contract number DE-AC02-98CH10886.